\documentclass[journal=jpccck,manuscript=article,layout=traditional]{achemso}
\usepackage{chemformula} % Formula subscripts using \ch{}
\usepackage[T1]{fontenc} % Use modern font encodings
\usepackage{upgreek}
\usepackage[utf8]{inputenc}
\author{Nathaniel Orndorf}
\author{Saranshu Singla}
\author{Ali Dhinojwala}
\email{ali4@uakron.edu}
\affiliation[Unknown University]
{Department of Polymer Science, The University of Akron, Akron, Ohio}

\title[An \textsf{achemso} demo]
  {Transition in the Acid-Base Component of Surface Free Energy of Ice upon the Premelting of its Second Molecular Bilayer}

\abbreviations{SFG, JKR, PDMS}
\keywords{Ice, Surface Premelting, Surface Energy, Acid-Base Interactions}

\begin{document}

\begin{tocentry}

 \includegraphics[width=1\columnwidth]{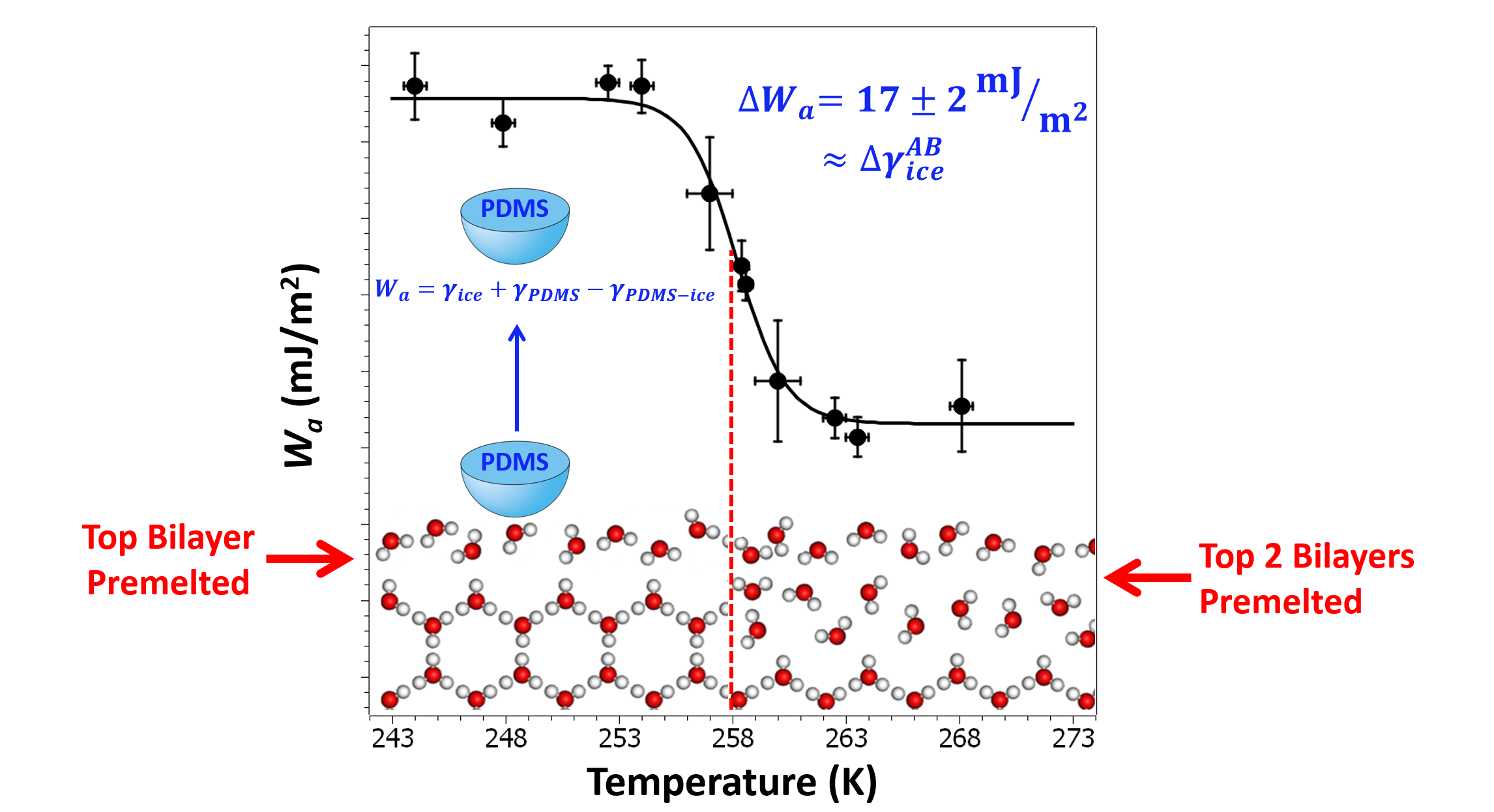}

\end{tocentry}
\begin{abstract}
Molecular disordering of the ice surface occurs below the bulk melting temperature of 273 K, termed surface premelting. The top-most molecular layer begins gradually premelting at 200 K, and has been linked to its low coefficient of friction through an increase in molecular mobility. The second molecular bilayer premelts around 257 K, but no study has linked this transition to a change in any macroscopic phenomena. Here, we show that the thermodynamic work of adhesion between polydimethylsiloxane (PDMS) and ice changes abruptly at $257.0\pm0.1$ K. Surface-sensitive sum frequency generation spectroscopy shows that there are no molecular level changes at the PDMS surface or the ice-PDMS interface near the transition in adhesion, indicating that the transition arises from changes of the ice surface. Using existing contact angle data in the literature, we show that this transition is due to a decrease in the acid-base component of the surface free energy of ice by $17\pm2$ mJ/m$^2$ at $257.0\pm0.1$ K. The change in surface energy provides a possible explanation for a variety of unexplained phenomena seen across the literature including ice adhesion, friction, and the morphology of snowflakes.
\end{abstract}


\section{Introduction}

Understanding how the molecular structure at the surface of solids dictates macroscopic properties such as surface free energy, adhesion, and friction has been a long standing problem in interfacial science and has applications in many areas. One particular area of high interest, but of which little is understood, is that of the surface of ice. Detailed knowledge on the surface of ice is important at all length scales, from antifreeze proteins keeping organisms alive well below freezing \cite{Voets2017} to sliding glaciers forming landscapes \cite{Persson2018}. The properties of ice are important in understanding the history of our planet \cite{Trinks2005} and the solar system \cite{Li2018b,Dundas2018}. From an application perspective, ice accretion and adhesion to aircraft \cite{Cao2018a}, wind turbines \cite{Afzal2018}, solar panels \cite{Andenaes2018}, and power lines \cite{Tomaszewski2019} decreases energy efficiency and presents dangerous situations. It is likely that climate change will cause icing events to shift in latitude, showing potential for these problems to become common in new locations \cite{Lambert2011}. Understanding friction on ice has implications in the grip of winter automobile tires and shoe soles \cite{Lahayne2016, Tuononen2016}. Unraveling the mechanism behind antifreeze proteins will advance technologies in agriculture and food preservation \cite{Voets2017}. In all of these examples, it is crucial to understand the surface and interfacial interactions of ice, both in terms of understanding our universe and advancing new materials and technologies that will increase the standard of living for people worldwide.

The surface of ice, or ice-air interface, has been increasingly studied since Michael Faraday first postulated its surface premelting in 1850 \cite{Faraday1850}. Since then, it has commonly been shown by several different techniques that the surface of ice behaves ‘liquid-like’ below the bulk melting temperature of 273 K \cite{Slater2019}. The technique most sensitive to surfaces and interfaces is sum frequency generation spectroscopy (SFG), which was first introduced in practice by Zhu \textit{et al.} in 1987 \cite{Zhu1987}. By overlapping a beam with a fixed frequency in the visible range $\omega_{vis}$ and a beam with a tunable frequency in the infrared range $\omega_{IR}$, the resulting SFG intensity at the sum of the beams’ frequency $\omega_{SFG}=\omega_{vis}+\omega_{IR}$ provides information on the ordering of molecules and the strength of interactions to depth of a few molecular layers \cite{Lambert2005, Shen2012}. SFG has been used to identify two specific premelting transitions of the ice surface. In 2001, Wei \textit{et al.} showed that the free O--H of the topmost molecular layer becomes increasingly disordered as the temperature is increased from 200 K \cite{Wei2001,Wei2002}. More recently, Sanchez \textit{et al.} showed that there is a sharp transition in the bonded O--H peak location around 257 K in both experiments and molecular simulations, which was attributed to the premelting of the second molecular bilayer \cite{AlejandraSanchez2017}. In the same study, they also confirmed that the intensity of the free O--H peak decreases with increasing temperature, but there is no shift in the free O--H peak location, showing that this transition has no major structural effects on the topmost layer. The molecular scale sensitivity of SFG provides detection of premelting on much smaller length scales (0.4 and 0.8 nm for the premelting of the first and second bilayer, respectively) than other techniques. Recent reviews give a more detailed history of ice studied by SFG \cite{Yamaguchi2019, Tang2020}. 

Such studies are being used to begin to understand the mechanism behind ice friction \cite{Weber2018}. Likewise, an attempt to link premelting to the shear adhesion strength of ice demonstrated that the premelt layer facilitates sliding \cite{Liljeblad2017}. But, through a collection of data across the literature, it has been shown that mechanical ice adhesion tests have large uncertainties across several orders of magnitude \cite{Work2018a}. This disagreement is likely due to energy dissipation in deforming the materials, rendering mechanical adhesion tests imprecise when aiming to extract thermodynamic quantities. These studies link the gradual disordering of the topmost molecular layer of the ice surface to gradual changes in the mechanical mobility of the surface molecules, resulting in low friction and sliding resistance \cite{Nagata2019}. However, a link between the precise premelting of the second molecular bilayer and surface properties has yet to be shown. Any reconfiguration of the molecules on the surface of ice likely results in a change in its surface free energy. The surface energy of medium $i$, $\gamma^{tot}_i$, can be written as a sum of energies of different interactions, namely the Lifshitz-van der Waals (LW) and acid-base (AB, which includes hydrogen bonding) interactions \cite{Fowkes1962}
\begin{equation}
\gamma_i^{tot}=\gamma_i^{LW}+\gamma_i^{AB}.
\label{SE}
\end{equation}
The surface energy of ice has been measured by contact angles of various liquids on ice \cite{Adamson1970, Kloubek1974, VanOss1992}, the contact angle of water at grain boundaries \cite{Ketcham1969}, homogeneous nucleation rates \cite{Boinovich2014}, and molecular simulations \cite{Qiu2018}. These studies generally result in $\gamma^{tot}_{ice}\sim100$ mJ/m$^2$, but precise determination of the surface free energy is difficult. Furthermore, no study has adequately examined the surface energy of ice over its premelting transitions. In 1970, Adamson \textit{et al.} measured the contact angle of carbon disulfide on ice at several temperatures in the range of 225 to 270 K, but no transition was found \cite{Adamson1970}. Nearly two decades later, van Oss \textit{et al.} used various probe liquids to measure $\gamma_{ice}^{AB}$ to be 60.4 and 48.0 mJ/m$^2$ at 253 and 265 K respectively, but they assumed a linear trend in $\gamma_{ice}^{AB}$ in order to extrapolate their results to 273 K \cite{VanOss1992}. 

Analogous to using the contact angle made between a liquid and a solid to study the surface energy of the solid, Johnson, Kendall, and Roberts’s theory of elastic adhesion (JKR) \cite{Johnson1971,Chaudhury1992} can be used to measure the thermodynamic work of adhesion between two solids. The thermodynamic work of adhesion is defined as the amount of reversible energy needed to separate an interface between media $i$ and $j$ and create two surfaces, expressed by the Dupr\'{e} Equation 
\begin{equation}
W_a=\gamma^{tot}_i+\gamma^{tot}_j-\gamma^{tot}_{ij},
\label{W}
\end{equation}
where $\gamma^{tot}_{ij}$ is the total interfacial free energy between media $i$ and $j$ \cite{Dupre1869}. Thus, measuring the work of adhesion between ice and another material allows one to make inferences on the surface energy of ice, provided that the surface energy and surface properties of the other material are known. 

In 1981, Roberts and Richardson conducted JKR ‘touch on’ experiments with a 1.8 cm diameter polyisoprene hemisphere and ice \cite{Roberts1981b}. They saw a drastic increase in the contact radius as the temperature approached 273 K. Rather than a true thermodynamic increase in $W_a$, this was attributed to the formation of a capillary bridge of the premelt layer between the hemisphere and the ice--a claim which is supported by a drop in friction and pull-off adhesion at the same temperature. They also noticed circular marks left on the ice from where the contact had been, which were more apparent with increasing dwell time, load, or temperature. Due to the capillary bridging resulting in inaccurate measurements of $W_a$, their study does not allow for conclusions about the surface free energy of ice to be drawn.

In this study, we use the JKR theory of elastic adhesion \cite{Johnson1971} to measure the thermodynamic work of adhesion between polydimethylsiloxane (PDMS) and ice over a wide temperature range. PDMS is often the host material used in low ice adhesion materials and coatings because of its low elastic modulus and surface energy \cite{Chen2017,Zhuo2018,Golovin2019,Irajizad2019a}, but the specific molecular interactions between ice and PDMS have not been identified, making it an ideal material for this study. Surface-sensitive SFG spectra of the PDMS surface (PDMS-vacuum interface) and the ice-PDMS interface are collected to examine any molecular-level changes occurring within the studied temperature range. The work of adhesion measurements between ice and PDMS, in conjunction with the surface-sensitive and molecular-level spectroscopic observations, allow us to make inferences on changes in the surface free energy of ice as a function of temperature.

\section{Experimental Methods}

\subsection{Thermodynamic Work of Adhesion Measurements}

The JKR model was derived for the contact between two elastic spheres, but by taking one sphere in the limit of infinite radius and infinite elastic modulus, the well-known JKR equation expresses the radius of the contact circle $a$ between a hard, flat plane and an elastic sphere of radius $R$ and elastic modulus $E$ by
\begin{equation}
a^3=\frac{9R}{16E}\left[P+3\pi RW_a+\sqrt{6\pi RPW_a+(3\pi RW_a)^2}\right].
\label{JKR}
\end{equation}
In the simplifying case that the load $P$ between the sphere and plane is negligible,
\begin{equation}
a^3=\frac{27\pi}{8}\frac{R^2}{E}W_a.
\label{ZeroJKR}
\end{equation}
By measuring the contact radius $a$ formed between an elastic spherical cap of known radius $R$ and elastic modulus $E$ and a flat surface at zero load, the thermodynamic work of adhesion can be extracted using Equation \ref{ZeroJKR}.

\begin{figure}
 \includegraphics[width=1\columnwidth]{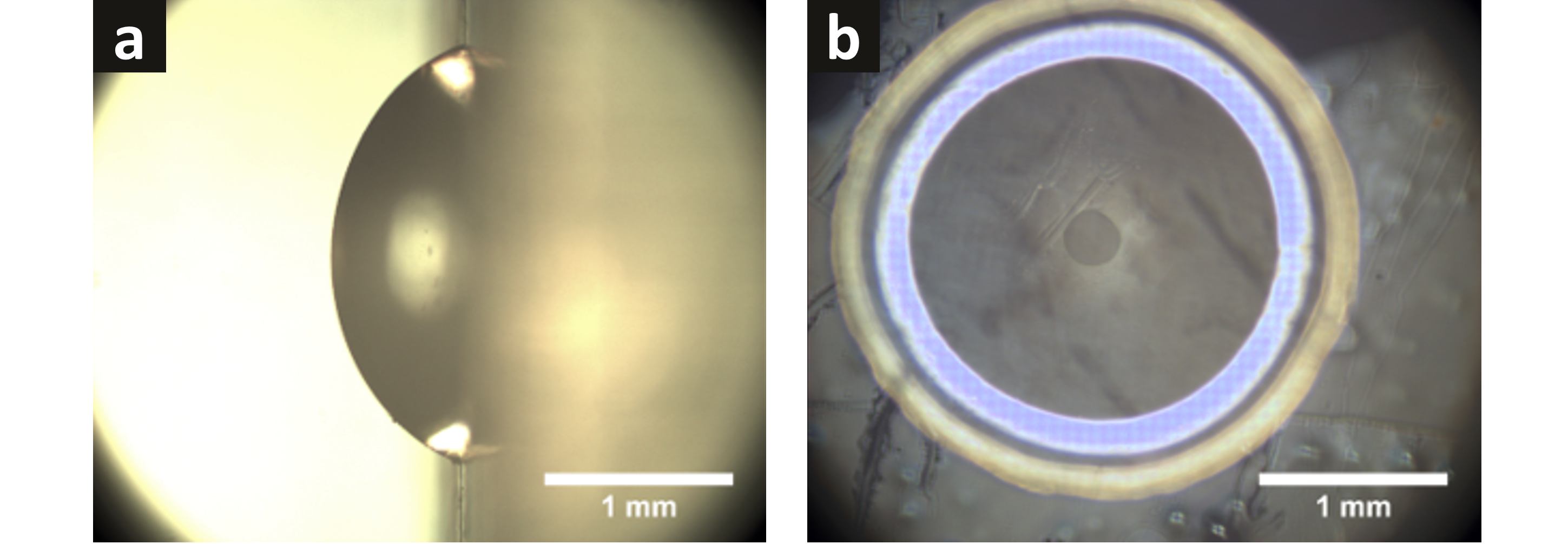}
 \caption{Side view of a PDMS spherical cap (a). When the spherical side is placed in contact with ice, the contact spot can clearly be seen by viewing through the flat side of the cap (b).\label{Caps}}
 \end{figure}

Smooth, spherical caps were fabricated using vinyl-terminated PDMS of $M_w=9,000$ g/mol (Gelest Inc. V-21). Tetrakis-dimethylsiloxysilane (SIT 7278.0) was added as a tetra-functional cross-linker at a 4.4 vinyl-to-hydride molar ratio to avoid unreacted species, resulting in a cross-linked molecular weight of $M_c=6000$ g/mol. The PDMS melt was placed as droplets on a fluorinated glass dish and cured at $60$ $^{\circ}$C for 3 days to form solid, soft spherical caps. The caps were Soxhlet-extracted with toluene at $124$ $^{\circ}$C for 24 hours, and then dried under vacuum. An optical microscope was used to capture side-view images of the caps shown in Figure \ref{Caps}, which were used to measure the radius $R$ ($\approx1$ mm) by fitting it to a circle. The elastic modulus of the caps was measured to be $E=1.9\pm0.1$ MPa by approach and retraction adhesion tests using the JKR theory. Further details of the PDMS cap fabrication and elastic modulus measurements are available in the literature \cite{Dalvi2019}. 

Purified water (Milipore Filtration System, $\rho=$18.2 M$\Omega\cdot$cm) was degassed using a sonicator, and poured in solvent cleaned ice cube trays. The trays were placed inside of a walk-in freezer at the specific temperature at which the experiments would take place. At least 24 hours after freezing, the ice cubes were removed from the tray and frozen onto aluminum plates for handling. To ensure a smooth surface, the ice was microtomed using cuts as small as 1 $\upmu$m thick. The crystal orientation at the surface was unknown, but, as discussed later in the text, our results are not expected to change on different crystal faces. Due to the degassing, the ice contained very few air bubbles. Bubbles, cracks, and grain boundaries were easily seen with an optical microscope, and those areas of the ice surface were avoided for the adhesion experiments.

All of the materials (including the PDMS caps and tweezers) used for the experiment were equilibrated inside of the walk-in freezer at the temperature of the experiment for at least 24 hours prior to the start of the experiment. The spherical side of a PDMS cap was carefully placed on the ice surface using tweezers, and the contact was viewed through the flat side of the cap using a microscope. After the contact area stabilized ($\approx1$ min), images were recorded (shown in Fig. \ref{Caps}), which were used to measure the contact radius $a$ to calculate $W_a$ using Equation \ref{ZeroJKR}. These measurements were repeated about five times at different locations on the ice in order to calculate an average and standard error for a single cap, and several caps of various radii were used. The experiment was replicated at several temperatures in the range of $243$ to $268$ K, using new PDMS caps and newly frozen ice for each experiment. The cap radius $R$ was measured inside of the walk-in freezer at each temperature that the experiments were conducted. Because the elastic modulus of the PDMS was measured at room temperature, the elastic modulus at the lower temperatures was calculated by scaling according to the affine model, given by $\frac{E(T_1)}{E(T_2)}=\frac{T_1}{T_2}$ \cite{Rubinstein2003}.

\subsection{Sum Frequency Generation Spectroscopy}

For SFG, 200 nm thick PDMS films were made by spin-coating a 4 wt.\% solution of vinyl terminated PDMS (Gelest Inc. V-31) of $M_w=28000$ in Hexane (Sigma-Aldrich, $\geq99\%$ purity) onto sapphire prisms for 70 s at 2000 rpm. The films were annealed at $60$ $^{\circ}$C for at least 6 hours and allowed to cool to room temperature before the spectroscopy experiments. The surface structure of the chemically cross-linked PDMS caps and the PDMS films which are not chemically cross-linked is the same because the CH$_3$ groups present at the surface do not change.

Our SFG setup consists of a Spectra Physics Laser system with a visible beam with a constant wavelength of 797 nm ($\approx70$ $\upmu$J energy, 1 ps pulse width, 1 kHz repetition rate, 1 mm diameter) and an IR beam with a tunable wavelength ($\approx3.5$ $\upmu$J energy, 1 ps pulse width, 1 kHz repetition rate, 100-200 $\upmu$m diameter). A total internal reflection geometry was used to probe the SFG signal generated at the surface of the previously described PDMS films on sapphire prisms at incident angles of 16$^{\circ}$ for the ice-PDMS interface and 42$^{\circ}$ for the PDMS surface. A custom built cooling stage allows for the collection of SFG spectra at precise temperatures. Because the cooling stage necessitates a vacuum around the sealed cell containing the prism, only PDMS-vacuum spectra can be taken below room temperature. Spectra of the PDMS-air and PDMS-vacuum interfaces show no detectable differences at room temperature (Figure S1). Spectra of the PDMS surface (PDMS-vacuum interface) were taken at various temperatures at 5 K intervals in the SSP polarization (s-polarized SFG output, s-polarized visible input, and p-polarized IR input). An equilibration time of at least 5 min at each temperature was provided before collecting each spectrum. To collect the ice-PDMS spectra, purified water (Milipore Filtration System, $\rho=$18.2 M$\Omega\cdot$cm) was sealed inside the cell and frozen next to the PDMS film at a rate of 0.3 K min$^{-1}$. Afterwards, the system was equilibrated for at least 5 min at each temperature before the SFG spectra were collected with PPP polarization (because SSP had low counts). Previous work has shown that the IR beam energy does not induce melting of the ice \cite{Zhang2014}. Further details on the SFG setup and cooling stage are available in the literature\cite{AnimDanso2013}.

All SFG spectra reported in this study were fit with Lorentzian peak functions
\begin{equation}
I_{SFG}\propto\left|\chi_{NR}+\sum\frac{A_q}{\omega_{IR}-\omega_q-i\Gamma_q}\right|^2,
\label{SFGeq}
\end{equation}
where $A_q$ and $\Gamma_q$ are the amplitude and damping constant of the $q$th vibrational frequency (peak center) $\omega_q$ with a nonresonant contribution $\chi_{NR}$ \cite{Lambert2005, Shen2012}. Only the peak centers are used for analysis in this study, but the remaining fit parameters are given in the Supporting Information. The first moment of a peak provides a weighted average of its center, and is calculated by the same method as previous studies \cite{AlejandraSanchez2017}.

\section{Results and Discussion}

\subsection{Thermodynamic Work of Adhesion}

\begin{figure}
 \includegraphics[width=1\columnwidth]{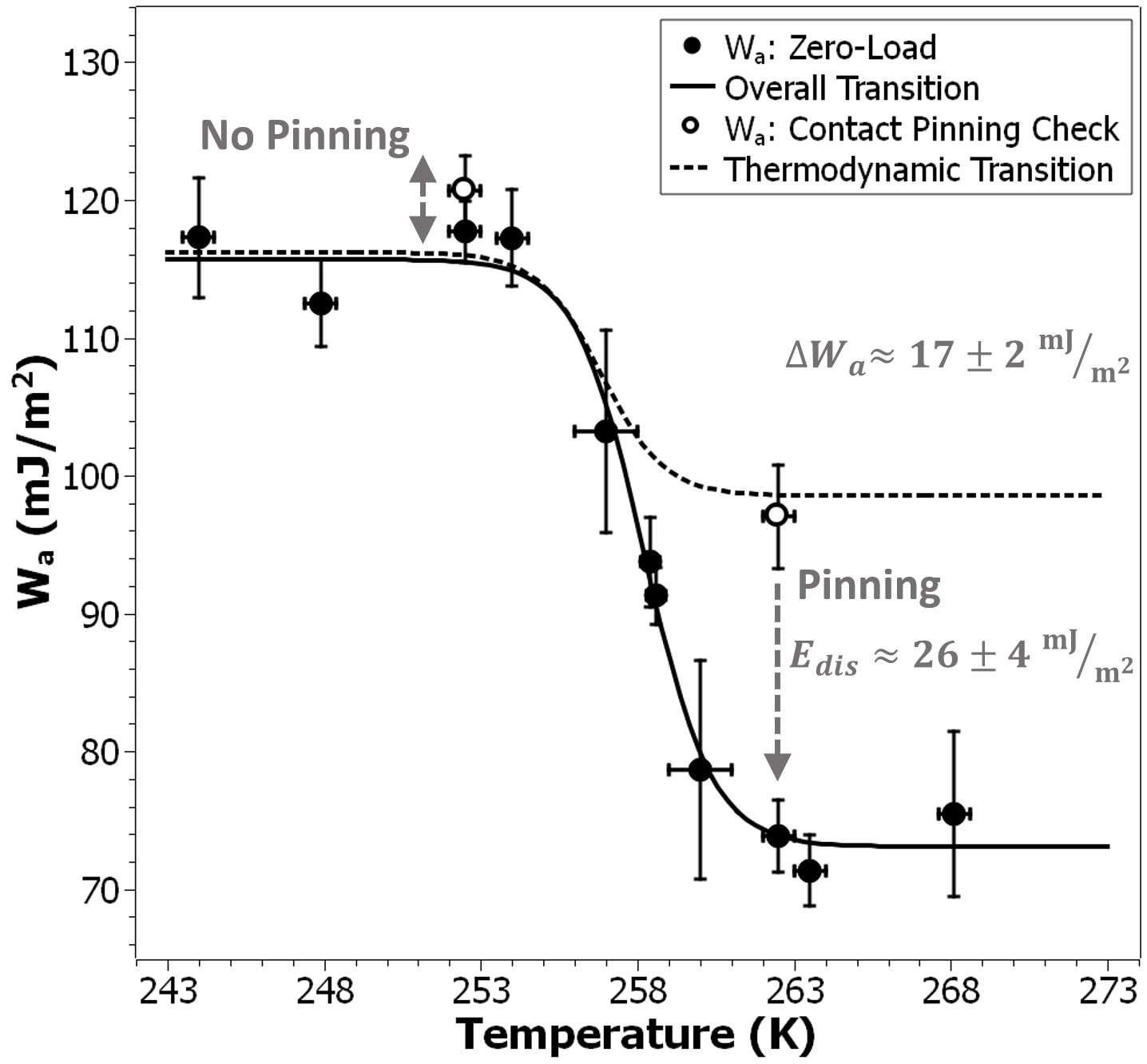}
 \caption{Measured work of adhesion between PDMS and ice. Both lines are sigmoidal fits using the logistic function.\label{Adhesion}}
 \end{figure}

\begin{figure}
 \includegraphics[width=1\columnwidth]{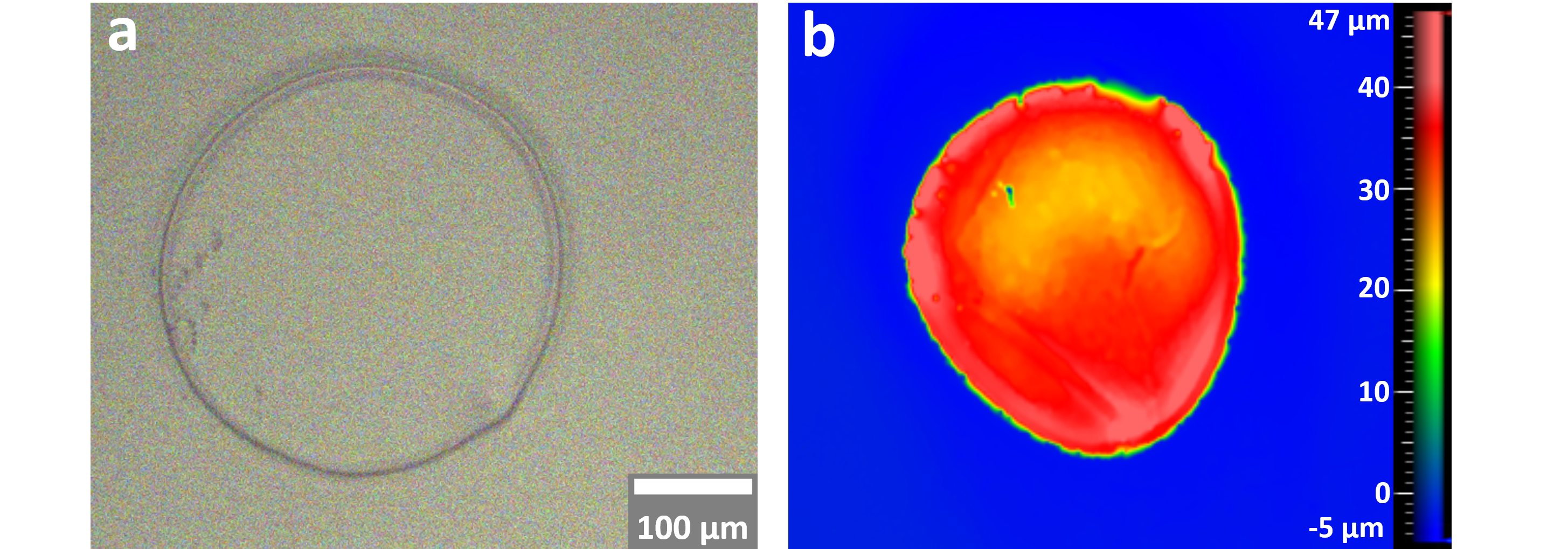}
 \caption{A mark the same shape as the contact area left on the ice surface (a). Optical profilometry of a surface replica (opposite in height) from a different mark confirms that the marks are impressions into the ice surface (b). \label{Impression}}
 \end{figure}

Figure \ref{Adhesion} displays the measured work of adhesion between PDMS and ice at various temperatures. A sigmoidal fit shows a transition from $116 \pm 2$  to $73 \pm 3$ mJ/m$^2$ when heated above $258.4 \pm 0.2$ K, which is interestingly similar to the temperature at which the second molecular bilayer of the ice surface premelts \cite{AlejandraSanchez2017}. When the small weight of the caps $\approx5$ mN is taken into account, Equation \ref{JKR} calculates $\approx5$ mJ/m$^2$ lower than the zero-load approximation shown in Figure \ref{Adhesion}, but the difference in the measurements before and after the transition is identical. Similar to Roberts and Richardson, we noticed circular marks left on the ice after longer dwell times, much more readily at temperatures above 258 K. Optical profilometer measurements of replicas of the ice surface (opposite in height) shown in Figure \ref{Impression} confirm that these circular marks are impressions left in the ice surface in the shape of the contact area. The only temperature for which these impressions occurred during the time frame of a measurement ($\approx1$ min) was 268 K, resulting in a larger apparent contact than the value of interest. The width of the impression circle on the surface of ice is measured to be $\approx9$ $\upmu$m, which is subtracted from the measured contact radii to calculate the true contact radius at 268 K. The presence of the impressions would introduce complications during a continuous temperature scan, so only independent measurements at different temperatures are studied here. The transition in the measured $W_a$ can arise from a thermodynamic change in $W_a$, from contact line pinning at some temperatures, or a combination of both.

To check for contact line pinning, we applied a higher load to the cap while placing it on the ice with the tweezers, released the load, and measured the contact area after it returned to stability. Repeating the same loading on a simple laboratory scale resulted in a highly variable applied load between 10 and 50 mN. Using this technique at 253 K, the contact area returns to what it was at zero-load, showing that there is no pinning below 258 K. Thus, $W_a=116\pm2$ mJ/m$^2$ below 258 K is expected to be a true thermodynamic value. Conversely, after applying a higher load at 263 K, the contact area does not return to that of the previous zero-load experiments. Instead, the contact area returns to the value corresponding to $99$ mJ/m$^2$. This shows that above 258 K, the true thermodynamic value $W_a\leq99$ mJ/m$^2$. While there was a high variance in the applied load, the contact area repeatedly returned to the same value, suggesting that the true thermodynamic value is around 99 mJ/m$^2$ above 258 K. Thus, we conclude that the remainder of the change arises from thermodynamic effects. To explore this thermodynamic change, we fit only the thermodynamic data with a sigmoidal curve, which shows a transition from $116\pm1$ to $99\pm2$ mJ/m$^2$ when heated above $257.0\pm0.1$ K (also shown in Fig. \ref{Adhesion}). This thermodynamic transition lies closer to the second molecular bilayer premelting transition temperature of ice, and consists of a difference in $W_a$ before and after the transition of $\Delta W_a=17\pm2$ mJ/m$^2$. 

\subsection{Sum Frequency Generation Spectroscopy}

\begin{figure}
\includegraphics[width=0.8\columnwidth]{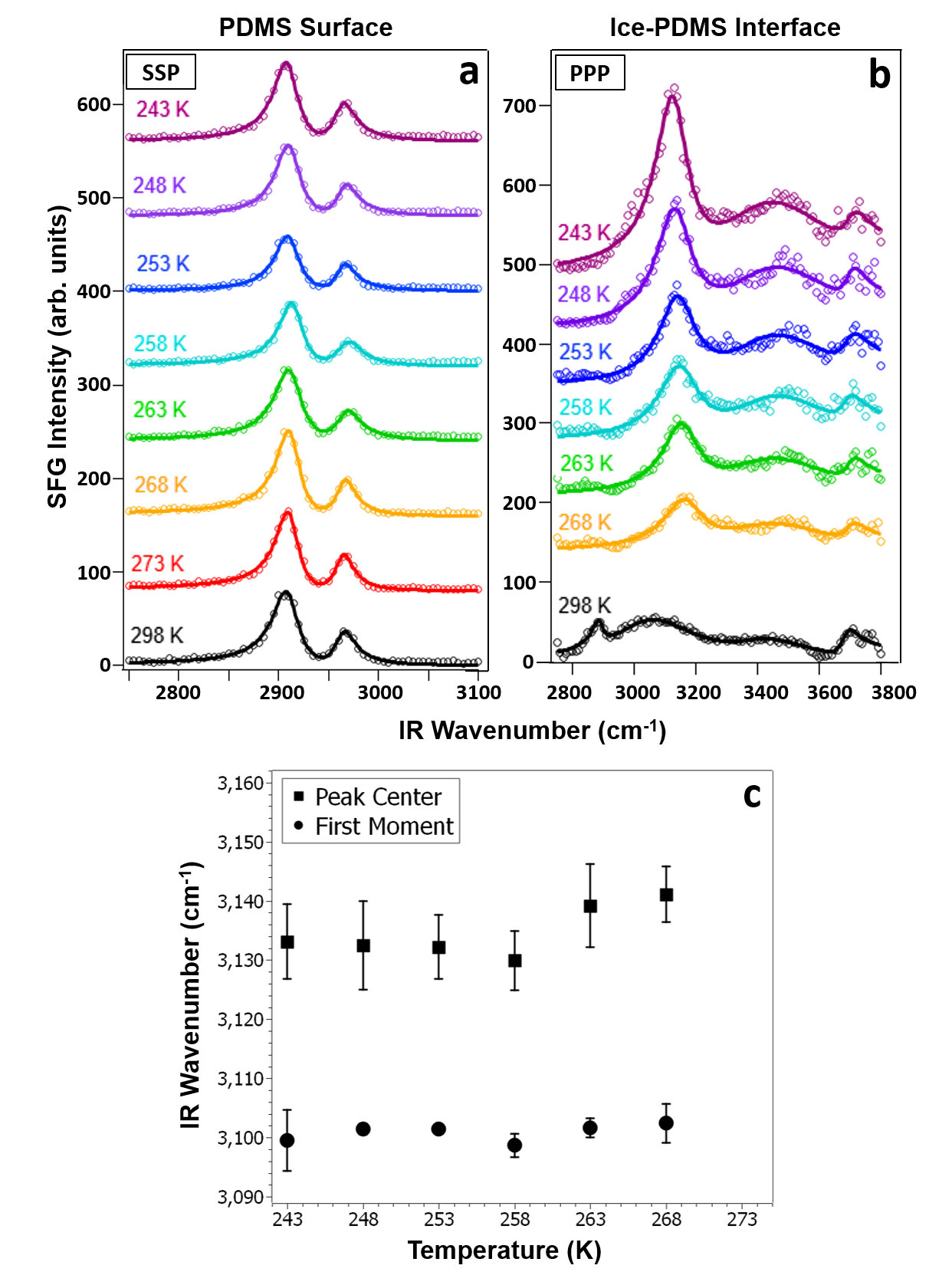}
\caption{Representative SFG spectra of the PDMS surface (a) and the ice-PDMS interface (b) at various temperatures. The spectra are offset for visual clarity, but are not scaled. The solid lines are Lorentzian peak fits. The ice-PDMS interface spectra show no shift in the center or first moment of the bonded O--H peak location (c). The error bars represent the variability between spectra of different samples. PPP polarized spectra of the PDMS surface and SSP polarized spectra of the ice-PDMS interface are given in the Supporting Information.  \label{SFG}}
\end{figure}

The SFG spectra of the PDMS surface (PDMS-vacuum interface) and ice-PDMS interface, and their corresponding fits, are shown in Figure \ref{SFG}. It is evident that the SFG spectra of the PDMS surface do not change in this temperature range. The peaks at $2911.5\pm0.7$ and $2965.4\pm0.6$ cm$^{-1}$ are assigned to the symmetric and asymmetric CH$_3$ stretch modes, respectively \cite{Yurdumakan2005, Kim2008, Nanjundiah2009, Shi2009}. The 2910 cm$^{-1}$ PDMS peak is present against water but not ice, which could be due to chemical interactions between the CH$_3$ groups and ice or simply due to the more intense SFG peaks from ice overshadowing those of PDMS. The ice-PDMS spectra show three peaks: a peak at $3135\pm3$ cm$^{-1}$ corresponding to the O--H stretch mode bonded in the ice lattice (four hydrogen bonds) \cite{Wei2001, Wei2002}, a broad peak centered at $3410\pm6$ cm$^{-1}$ corresponding to lower coordination hydrogen bonds (less than four) \cite{AnimDanso2013}, and a peak at $3708\pm17$ cm$^{-1}$ corresponding to the free O--H stretch mode \cite{Wei2001, Wei2002}. A signal from the sapphire free O--H at the buried sapphire-PDMS interface produces noise and higher uncertainty in the 3700 cm$^{-1}$ peak \cite{Defante2015}. The $3135$ cm$^{-1}$ peak appears at a lower wavenumber than that of the ice surface \cite{Wei2001,Wei2002,AlejandraSanchez2017}, indicating that the hydrogen bond interactions between water molecules may be stronger against the PDMS interface than against the air interface. Unlike the ice surface \cite{AlejandraSanchez2017}, this peak shows no sudden shift in either the peak center or the first moment with a change in temperature between 243 and 268 K. No sudden shift in this peak indicates that there is no premelting transition in this temperature range. It could be the case that the premelting of the second bilayer is shifted to either higher or lower temperatures. In either case, it is evident that the transition in the work of adhesion does not arise from structural changes on the PDMS surface or the ice-PDMS interface. Thus, by reasoning analogous to Equation \ref{W}, we conclude that the transition in the work of adhesion is due to changes on the ice surface, specifically at the same temperature at which the second molecular bilayer premelts.  

\subsection{Surface Free Energy of Ice}

Because our SFG results show that there are no molecular-level changes in the PDMS surface or the ice-PDMS interface, we expect $\gamma^{tot}_{PDMS}$ and $\gamma^{tot}_{ice-PDMS}$ to vary linearly over this temperature range. As the transition $\Delta W_a=\Delta\gamma^{tot}_{ice}+\Delta\gamma^{tot}_{PDMS}-\Delta\gamma^{tot}_{ice-PDMS}$ is sudden (i.e. $\Delta$ represents a change over a few K or a discontinuity), the linear trends in $\gamma^{tot}_{PDMS}$ and $\gamma^{tot}_{ice-PDMS}$ result in $\Delta\gamma^{tot}_{PDMS}\approx\Delta\gamma^{tot}_{ice-PDMS}\approx0$, and do not contribute to $\Delta W_a=17\pm2$ mJ/m$^2$. Thus, $\Delta W_a\approx\Delta\gamma^{tot}_{ice}=\Delta\gamma^{LW}_{ice}+\Delta\gamma^{AB}_{ice}$. Adamson \textit{et al.} have shown that the contact angle of carbon disulfide on ice changes gradually and linearly over this temperature range \cite{Adamson1970}. Because carbon disulfide is capable of only dispersive interactions $(\gamma^{tot}_{CS_2}=\gamma^{LW}_{CS_2})$, this shows that $\gamma^{LW}_{ice}$ varies linearly over the transition, so $\Delta\gamma^{LW}_{ice}\approx0$. Thus, we conclude that $\Delta W_a\approx\Delta\gamma^{AB}_{ice}\approx17\pm2$ mJ/m$^2$, meaning that the acid base component of the surface free energy of ice abruptly decreases by $17\pm2$ mJ/m$^2$ at $257.0\pm0.1$ K, the same temperature at which the second molecular bilayer of the surface of ice premelts \cite{AlejandraSanchez2017}. This value is similar to van Oss \textit{et al.}'s measurement of $\Delta\gamma_{ice}^{AB}=12.4$ mJ/m$^2$ between two measurements at 253 and 265 K \cite{VanOss1992}, but they suggested a linear decrease in $\gamma_{ice}^{AB}$ rather than a transition at 257 K. Because the second molecular bilayer premelts at slightly different temperatures on different crystal faces \cite{AlejandraSanchez2017}, we expect $\gamma_{ice}^{AB}$ of different faces to transition at slightly different temperatures. The difference in these temperatures is so close (1.7 K) that the methods used in this study are unlikely to be able to detect a difference in the $\gamma_{ice}^{AB}$ transition temperature.

Until now, the premelting of ice has been linked to macroscopic phenomena only through the gradual increase in mobility of the top-most surface molecules with increasing temperature. Our results show that there is also a change in the acid-base component of the surface free energy of ice at the same temperature at which the second molecular bilayer premelts. This transition has ramifications in many areas. With a sudden drop in $\gamma_{ice}$ above 257 K, we expect the work of adhesion between ice and any material to also show a sudden drop above 257 K, provided that $\Delta\gamma_{ice}>\Delta\gamma_{ice-material}$. Roberts and Richardson saw a drastic decrease in both friction and pull-off adhesion force of polyisoprene on ice \cite{Roberts1981b}, both of which can be explained by the drop in surface energy of ice around 257 K. A similar transition seems plausible in the fracture toughness of ice \cite{Nixon1987}. The structure and surface energy of the premelt layer of ice is expected to have an influence on crystal growth from the vapor phase, which would influence the morphology of snowflakes \cite{Libbrecht}. Interestingly, the morphologies of snowflakes transition from long, slender columns to large, thin plates when the temperature reaches about 258 K \cite{Libbrecht2005}, possibly due to $\gamma_{ice}^{AB}$ transitioning at slightly different temperatures on different crystal faces.

\section{Conclusion}

In summary, we have shown that the thermodynamic work of adhesion between PDMS and ice drastically transitions at 257 K, the temperature at which the second molecular bilayer of the ice surface premelts. Unlike the ice-air interface, SFG spectra of both the PDMS surface and ice-PDMS interface show no sharp transition around 257 K, confirming that the transition in adhesion is due to a transition of the ice surface. Using contact angle data of carbon disulfide (capable of only dispersive interactions) on ice from the literature, we have shown that it is the acid-base component of the ice surface which shows a transition of $\Delta\gamma_{ice}^{AB}\approx17\pm2$ mJ/m$^2$. There is also pinning of the advancing contact line above 258 K, possibly due to energy dissipation in the mobility of the premelted bilayers. Understanding how the molecular structure of the premelt layer dictates the surface energy of ice has direct implications on many scientific fields, from physicists aiming to understand adhesion and friction to biologists unraveling the mechanisms behind antifreeze proteins.
\begin{acknowledgement}

We thank Siddhesh Dalvi for the fabrication of the PDMS caps and Edward Laughlin for the design and construction of the SFG cooling stage. Insightful comments on the results were graciously provided by Manoj Chaudhury and Michael C. Wilson. We also thank Mario Vargas and Andrew Work for discussions on the overall topic. Funding was provided by the National Science Foundation (NSF DMR-1610483) and the project was partially funded by a Goodyear Tire and Rubber Company fellowship. 

\end{acknowledgement}
\begin{suppinfo}
\begin{itemize}
  \item SFG spectra of the PDMS surface in the PPP polarization as a function of temperature,
  \item SFG Spectra of the ice-PDMS interface in the SSP polarization as a function of temperature, and
  \item fit parameters for all SFG spectra shown in the text and Supporting Information.
\end{itemize}
\end{suppinfo}
%\bibliography{library}

\providecommand{\latin}[1]{#1}
\makeatletter
\providecommand{\doi}
  {\begingroup\let\do\@makeother\dospecials
  \catcode`\{=1 \catcode`\}=2 \doi@aux}
\providecommand{\doi@aux}[1]{\endgroup\texttt{#1}}
\makeatother
\providecommand*\mcitethebibliography{\thebibliography}
\csname @ifundefined\endcsname{endmcitethebibliography}
  {\let\endmcitethebibliography\endthebibliography}{}

\end{document}


%\begin{tocentry}
%\end{tocentry}


%\begin{abstract}
%\end{abstract}


\section{Additional SFG Spectra}

\begin{figure}
\includegraphics[width=0.6\columnwidth]{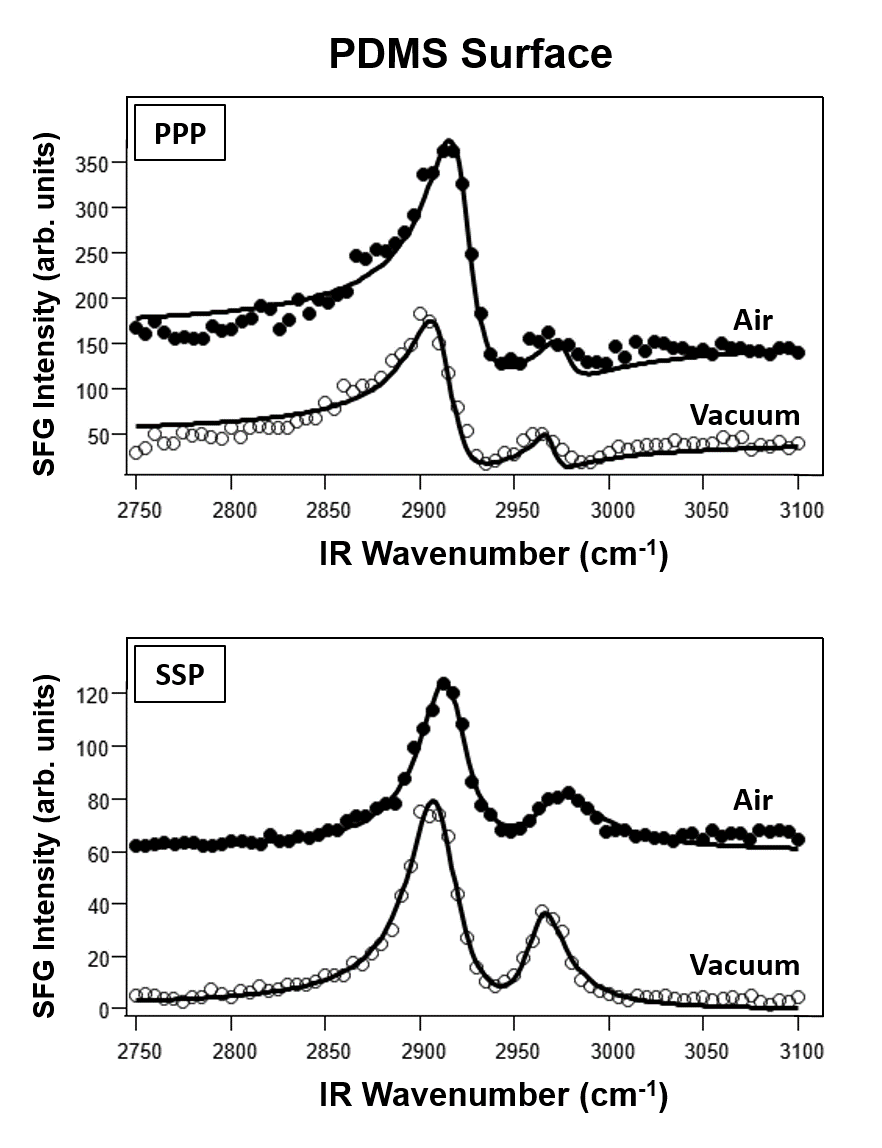}
\caption{SFG spectra of the PDMS-air and PDMS-vacuum interfaces at 298 K in the PPP and SSP polarizations. The spectra are offset for visual clarity, but are not scaled. The solid lines are Lorentzian peak fits.\label{Air}}
\end{figure}

\begin{figure}
\includegraphics[width=0.6\columnwidth]{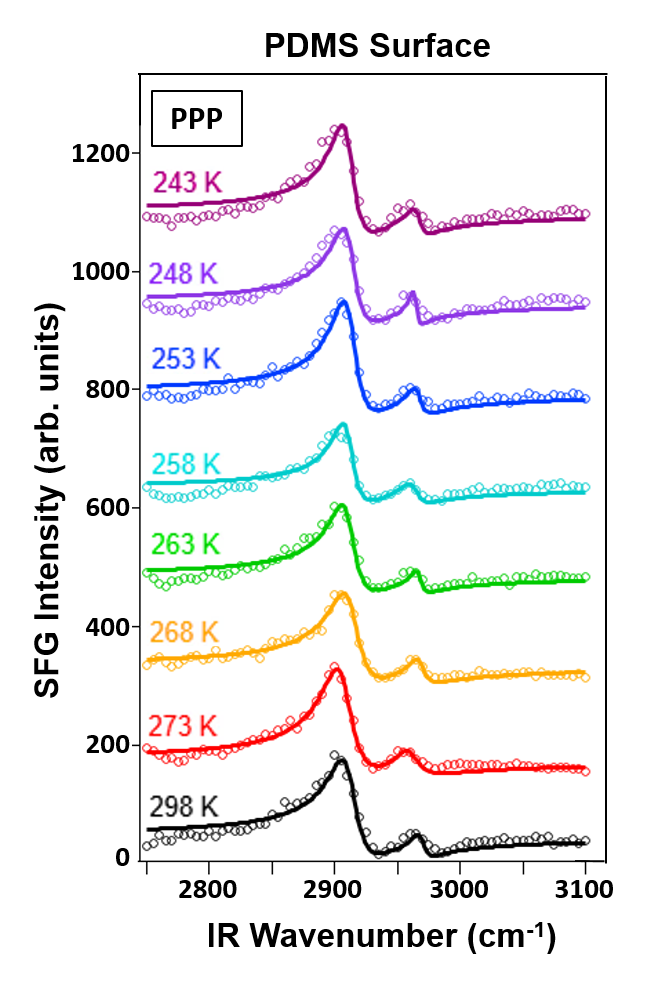}
\caption{SFG spectra of the PDMS surface (PDMS-vacuum interface) in the PPP polarization. The spectra are offset for visual clarity, but are not scaled. The solid lines are Lorentzian peak fits.\label{PDMSvacPPP}}
\end{figure}

\begin{figure}
\includegraphics[width=0.6\columnwidth]{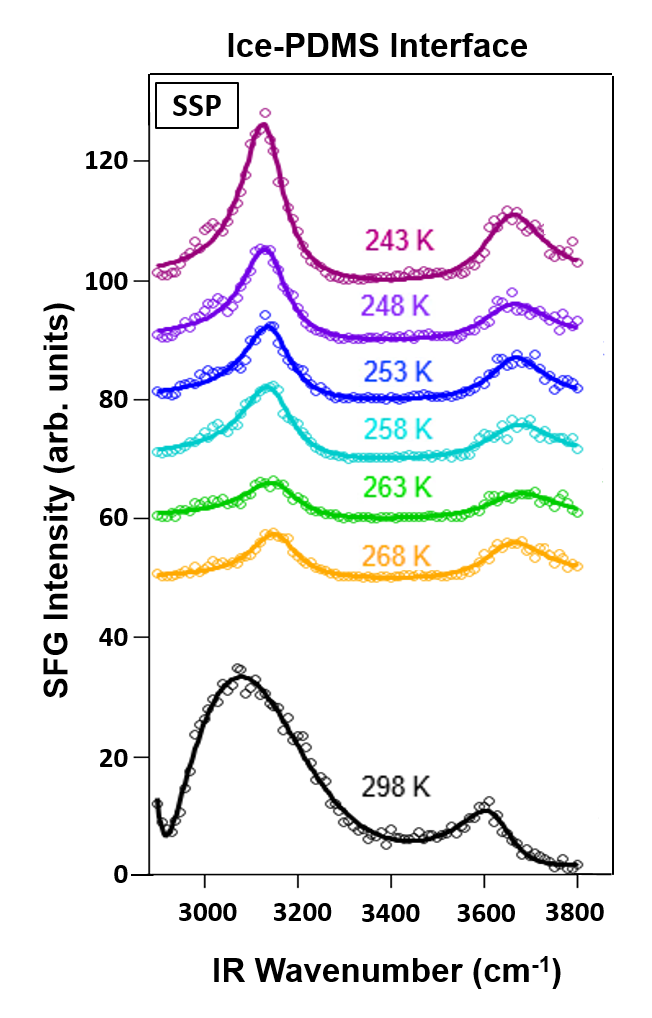}
\caption{SFG spectra of the ice-PDMS interface in the SSP polarization. The spectra are offset for visual clarity, but are not scaled. The solid lines are Lorentzian peak fits.\label{IcePDMSSSP}}
\end{figure}

\clearpage

\section{SFG Fitting}

All parameters in Equation 5 from the main text are allowed to vary freely in order to obtain the best empirical fit of the SFG spectra to the Lorentzian peak functions.

\begin{table}[]
\begin{tabular}{l|lllllll}
\multicolumn{1}{c|}{T (K)} & \multicolumn{1}{c}{$\omega_1$} & \multicolumn{1}{c}{$A_1$} & \multicolumn{1}{c}{$\Gamma_1$} & \multicolumn{1}{c}{$\omega_2$} & \multicolumn{1}{c}{$A_2$} & \multicolumn{1}{c}{$\Gamma_2$} & \multicolumn{1}{c}{$\chi_{NR}$} \\ \hline
243                        & 2909.4 $\pm$ 0.5                   & 130 $\pm$ 5                   & 15.0 $\pm$ 0.6                     & 2963.2 $\pm$ 0.8                   & 73 $\pm$ 6                    & 14 $\pm$ 1                         & -0.3 $\pm$ 0.1                      \\
248                        & 2911.3 $\pm$ 0.6                   & 126 $\pm$ 5                   & 15.2 $\pm$ 0.6                     & 2965.6 $\pm$ 0.8                   & 62 $\pm$ 6                    & 13 $\pm$ 1                         & -0.2 $\pm$ 0.1                      \\
253                        & 2910.5 $\pm$ 0.5                   & 110 $\pm$ 4                   & 15.0 $\pm$ 0.6                     & 2964.4 $\pm$ 0.7                   & 54 $\pm$ 5                    & 12 $\pm$ 1                         & -0.2 $\pm$ 0.1                      \\
258                        & 2915.2 $\pm$ 0.5                   & 110 $\pm$ 5                   & 14.5 $\pm$ 0.6                     & 2968 $\pm$ 1                       & 68 $\pm$ 6                    & 16 $\pm$ 2                         & -0.5 $\pm$ 0.1                      \\
263                        & 2912.3 $\pm$ 0.5                   & 136 $\pm$ 5                   & 16.7 $\pm$ 0.6                     & 2967.2 $\pm$ 0.8                   & 70 $\pm$ 6                    & 15 $\pm$ 1                         & -0.5 $\pm$ 0.1                      \\
268                        & 2912.5 $\pm$ 0.3                   & 128 $\pm$ 4                   & 14.5 $\pm$ 0.4                     & 2965.9 $\pm$ 0.6                   & 67 $\pm$ 4                    & 12 $\pm$ 0.8                       & -0.92 $\pm$ 0.09                    \\
273                        & 2911.5 $\pm$ 0.3                   & 110 $\pm$ 3                   & 14.1 $\pm$ 0.4                     & 2964.3 $\pm$ 0.5                   & 64 $\pm$ 3                    & 11.8 $\pm$ 0.7                     & -0.90 $\pm$ 0.08                    \\
298                        & 2909.7 $\pm$ 0.5                   & 134 $\pm$ 5                   & 16.1 $\pm$ 0.5                     & 2964.3 $\pm$ 0.7                   & 65 $\pm$ 5                    & 13 $\pm$ 1                         & -0.6 $\pm$ 0.1                     
\end{tabular}
\caption{Fitting parameters for the representative SSP PDMS-vacuum SFG spectra given in the main text.}
\label{pdmsssptab}
\end{table}

\begin{table}[]
\begin{tabular}{l|lllllll}
\multicolumn{1}{c|}{T (K)} & \multicolumn{1}{c}{$\omega_1$} & \multicolumn{1}{c}{$A_1$} & \multicolumn{1}{c}{$\Gamma_1$} & \multicolumn{1}{c}{$\omega_2$} & \multicolumn{1}{c}{$A_2$} & \multicolumn{1}{c}{$\Gamma_2$} & \multicolumn{1}{c}{$\chi_{NR}$} \\ \hline
243                        & 2911.5 $\pm$ 0.8                   & 109 $\pm$ 10                  & 11 $\pm$ 1                         & 2967 $\pm$ 3                       & 26 $\pm$ 10                   & 7 $\pm$ 3                          & -7.0 $\pm$ 0.2                      \\
248                        & 2912.7 $\pm$ 0.8                   & 104 $\pm$ 10                  & 11 $\pm$ 1                         & 2964 $\pm$ 1                       & 16 $\pm$ 8                    & 3 $\pm$ 4                          & -6.8 $\pm$ 0.2                      \\
253                        & 2912.8 $\pm$ 0.6                   & 121 $\pm$ 8                   & 11.5 $\pm$ 0.8                     & 2967 $\pm$ 2                       & 28 $\pm$ 8                    & 7 $\pm$ 2                          & -6.6 $\pm$ 0.1                      \\
258                        & 2911.5 $\pm$ 0.9                   & 86 $\pm$ 9                    & 10 $\pm$ 1                         & 2964.0 $\pm$ 0.6                   & 28 $\pm$ 7                    & 8 $\pm$ 3                          & -5.8 $\pm$ 0.2                      \\
263                        & 2911.5 $\pm$ 0.7                   & 111 $\pm$ 9                   & 12 $\pm$ 1                         & 2967.6 $\pm$ 2                     & 21 $\pm$ 8                    & 5 $\pm$ 3                          & -5.9 $\pm$ 0.1                      \\
268                        & 2913.8 $\pm$ 0.4                   & 128 $\pm$ 6                   & 13.9 $\pm$ 0.7                     & 2968 $\pm$ 1                       & 35 $\pm$ 6                    & 8 $\pm$ 2                          & -5.73 $\pm$ 0.09                    \\
273                        & 2907.4 $\pm$ 0.5                   & 157 $\pm$ 10                  & 14.6 $\pm$ 0.8                     & 2960 $\pm$ 2                       & 57 $\pm$ 10                   & 11 $\pm$ 3                         & -4.8 $\pm$ 0.1                      \\
298                        & 2912.7 $\pm$ 0.7                   & 122 $\pm$ 9                   & 13 $\pm$ 1                         & 2970 $\pm$ 2                       & 22 $\pm$ 8                    & 6 $\pm$ 3                          & -6.7 $\pm$ 0.1                     
\end{tabular}
\caption{Fitting parameters for the representative PPP PDMS-vacuum SFG spectra given in the Supporting Information.}
\label{pdmsppptab}
\end{table}

\begin{table}[]
\begin{tabular}{lllllll}
\multicolumn{1}{c|}{T (K)} & \multicolumn{1}{c}{$\omega_1 $} & \multicolumn{1}{c}{$A_1$}  & \multicolumn{1}{c}{$\Gamma_1 $} & \multicolumn{1}{c}{$\omega_2 $}  & \multicolumn{1}{c}{$A_2$} & \multicolumn{1}{c}{$\Gamma_2 $} \\ \hline
\multicolumn{1}{l|}{243}   & $3133\pm6$                      & $689\pm57$                 & 61 $\pm$ 4                          & $3413\pm9$                       & 1618 $\pm$ 70                & 212 $\pm$ 11                        \\
\multicolumn{1}{l|}{248}   & $3133 \pm 8    $                    & 657 $\pm$ 67                   & 70 $\pm$ 5                          & 3418 $\pm$ 8                         & 1598 $\pm$ 64                & 222 $\pm$ 22                        \\
\multicolumn{1}{l|}{253}   & $3132 \pm 5      $                  & 597 $\pm$ 40                   & 72 $\pm$ 4                          & 3421 $\pm$ 8                        & 1251 $\pm$ 62                & 210 $\pm$ 19                        \\
\multicolumn{1}{l|}{258}   & $3130 \pm 5    $                    & 586 $\pm$ 51                   & 58 $\pm$ 6                          & 3427 $\pm$ 9                        & 1196 $\pm$ 67                & 205 $\pm$ 17                        \\
\multicolumn{1}{l|}{263}   & $3139 \pm 7  $                      & 501 $\pm$ 68                   & 72 $\pm$ 6                          & 3416 $\pm$ 10                        & 1204 $\pm$ 74               & 217 $\pm$ 26                        \\
\multicolumn{1}{l|}{268}   & $3141 \pm 5$                        & 454 $\pm$ 63                   & 78 $\pm$ 8                          & 3418 $\pm$ 11                        & 1029 $\pm$ 58                & 218 $\pm$ 20                        \\
                           &                                 &                            &                                 &                                  &                           &                                 \\
                           & \multicolumn{1}{c}{$\omega_3 $} & \multicolumn{1}{c}{$A_3 $} & \multicolumn{1}{c}{$\Gamma_3 $} & \multicolumn{1}{c}{$\chi_{NR} $} &                           &                                 \\ \cline{2-5}
                           & $3704\pm9$                      & 106 $\pm$ 41                   & 44 $\pm$ 18                        & 1.1 $\pm$ 0.2                        &                           &                                 \\
                           & 3699 $\pm$ 6                        & 91 $\pm$ 29                    & 37 $\pm$ 11                         & 1.8 $\pm$ 0.3                        &                           &                                 \\
                           & 3697 $\pm$ 6                        & 119 $\pm$ 28                  & 46 $\pm$ 11                          & 1.8 $\pm$ 0.2                        &                           &                                 \\
                           & 3701 $\pm$ 4                        & 39 $\pm$ 14                    & 18 $\pm$ 7                          & 1.4 $\pm$ 0.3                        &                           &                                 \\
                           & 3703 $\pm$ 6                        & 85 $\pm$ 27                    & 38 $\pm$ 12                         & 1.2 $\pm$ 0.2                        &                           &                                 \\
                           & 3701 $\pm$ 5                      & 71 $\pm$ 18                    & 35 $\pm$ 10                         & 0.9 $\pm$ 0.2                        &                           &                                
\end{tabular}
\caption{Fitting parameters for the representative PPP ice-PDMS SFG spectra given in the main text.}
\label{icepdmsppp}
\end{table}

\begin{table}[]
\begin{tabular}{l|lllllll}
\multicolumn{1}{c|}{T (K)} & \multicolumn{1}{c}{$\omega_1 $} & \multicolumn{1}{c}{$A_1$} & \multicolumn{1}{c}{$\Gamma_1 $} & \multicolumn{1}{c}{$\omega_2 $} & \multicolumn{1}{c}{$A_2$} & \multicolumn{1}{c}{$\Gamma_2 $} & \multicolumn{1}{c}{$\chi_{NR} $} \\ \hline
243                        & 3131 $\pm$ 2                        & 274 $\pm$ 7                   & 55 $\pm$ 2                          & 3654 $\pm$ 5                        & 246 $\pm$ 14                  & 76 $\pm$ 5                          & -0.08 $\pm$ 0.08                     \\
248                        & 3136 $\pm$ 3                        & 219 $\pm$ 8                   & 58 $\pm$ 2                          & 3657 $\pm$ 7                        & 202 $\pm$ 15                  & 85 $\pm$ 8                          & -0.09 $\pm$ 0.08                     \\
253                        & 3144 $\pm$ 3                        & 202 $\pm$ 8                   & 60 $\pm$ 3                          & 3664 $\pm$ 5                        & 189 $\pm$ 11                  & 73 $\pm$ 5                          & -0.15 $\pm$ 0.08                     \\
258                        & 3144 $\pm$ 3                        & 216 $\pm$ 8                   & 65 $\pm$ 3                          & 3670 $\pm$ 6                        & 208 $\pm$ 13                  & 89 $\pm$ 7                          & -0.15 $\pm$ 0.08                     \\
263                        & 3155 $\pm$ 5                        & 163 $\pm$ 10                  & 70 $\pm$ 5                          & 3673 $\pm$ 8                        & 184 $\pm$ 13                  & 92 $\pm$ 8                          & -0.11 $\pm$ 0.09                     \\
268                        & 3155 $\pm$ 3                        & 159 $\pm$ 6                   & 60 $\pm$ 3                          & 3653 $\pm$ 4                        & 185 $\pm$ 9                   & 79 $\pm$ 5                          & 0.10 $\pm$ 0.06                     
\end{tabular}
\caption{Fitting parameters for the representative SSP ice-PDMS SFG spectra given in the Supporting Information.}
\label{icepdmsssp}
\end{table}

\begin{table}[]
\begin{tabular}{clllllll}
\multicolumn{1}{c|}{Polarization} & \multicolumn{1}{c}{$\omega_1 $} & \multicolumn{1}{c}{$A_1 $} & \multicolumn{1}{c}{$\Gamma_1 $} & \multicolumn{1}{c}{$\omega_2 $} & \multicolumn{1}{c}{$A_2$} & \multicolumn{1}{c}{$\Gamma_2 $} &                                  \\ \cline{1-7}
\multicolumn{1}{c|}{PPP}          & 2887 $\pm$ 3                        & 56 $\pm$ 13                    & 22 $\pm$ 5                          & 3045 $\pm$ 9                        & 1214 $\pm$ 88                 & 178 $\pm$ 12                        &                                  \\
\multicolumn{1}{c|}{SSP}          & \multicolumn{1}{c}{---}                        & \multicolumn{1}{c}{---}                   & \multicolumn{1}{c}{---}                         & 3111 $\pm$ 15                       & 1227 $\pm$ 125                & 235 $\pm$ 11                        &                                  \\
\multicolumn{1}{l}{}              &                                 &                            &                                 &                                 &                           &                                 &                                  \\
                                  & \multicolumn{1}{c}{$\omega_3$}  & \multicolumn{1}{c}{$A_3 $} & \multicolumn{1}{c}{$\Gamma_3 $} & \multicolumn{1}{c}{$\omega_4 $} & \multicolumn{1}{c}{$A_4$} & \multicolumn{1}{c}{$\Gamma_4 $} & \multicolumn{1}{c}{$\chi_{NR} $} \\ \cline{2-8} 
\multicolumn{1}{l}{}              & 3398 $\pm$ 12                       & 337 $\pm$ 116                  & 144 $\pm$ 31                        & 3684 $\pm$ 3                        & 107 $\pm$ 11                  & 33 $\pm$ 5                         & 1.3 $\pm$ 0.1                        \\
\multicolumn{1}{l}{}              & \multicolumn{1}{c}{---}         & \multicolumn{1}{c}{---}    & \multicolumn{1}{c}{---}         & 3626 $\pm$ 5                        & 141 $\pm$ 21                  & 72 $\pm$ 9                          & -3.8 $\pm$ 0.2                      
\end{tabular}
\caption{Fitting parameters for the representative PPP (main text) and SSP (Supporting Information) water-PDMS SFG spectra at 298 K.}
\label{waterpdms}
\end{table}

%\end{landscape}


%\begin{acknowledgement}
%\end{acknowledgement}


%\begin{suppinfo}
%\end{suppinfo}

\bibliography{achemso-demo}